\documentclass[prd,showpacs,reprint,nofootinbib,showkeys]{revtex4-1}

\usepackage[utf8]{inputenc}
\usepackage{amsmath}
\usepackage{amsfonts}
\usepackage{amssymb}
\usepackage{graphicx}
\usepackage{hyperref}
\usepackage{braket}
\usepackage{multirow}
\usepackage{enumitem}
\usepackage{bbm}

\newcommand{\Obs}{{\cal O}}

\def\sl3c{\text{SL}(3,\mathbb{C})}
\def\su3{\text{SU(3)}}

\begin{document}

\title{Reweighting complex Langevin trajectories}

\author{Jacques Bloch}
\email{jacques.bloch@ur.de}
\affiliation{Institute for Theoretical Physics, University of Regensburg, 93040
  Regensburg, Germany}

\date{March 6, 2017}

\begin{abstract}
Although the complex Langevin method can solve the sign problem in simulations of theories with complex actions, the method will yield the wrong results if known validity conditions are not satisfied. We present a novel method to compute observables for a target ensemble by reweighting complex trajectories generated with the complex Langevin method for an auxiliary ensemble having itself a complex action. While it is imperative that the validity conditions be satisfied for the auxiliary ensemble, there are no such requirements for the target ensemble. This allows us to enlarge the applicability range of the complex Langevin method. We illustrate this at the hand of a one-dimensional partition function and two-dimensional strong-coupling QCD.
\end{abstract}


\maketitle

\section{Introduction}

Lattice simulations of quantum chromodynamics (QCD) at nonzero chemical potential are especially challenging as the complex fermion determinant rules out the use of importance sampling methods. Current solutions to circumvent this sign problem have a computational cost which grows exponentially with the volume and are restricted to regions of parameter space which do not encompass the phenomenologically interesting phase transition \cite{Fodor:2015doa}. An alternative solution method that has attracted a lot of attention recently is the complex Langevin (CL) method \cite{Parisi:1984cs,Aarts:2013uxa}, which uses stochastic differential equations for complexified degrees of freedom to sample the partition function and compute expectation values. The hope that this method may be used to investigate the QCD phase transition has recently been dented by investigations in heavy-dense \cite{Sexty:2013ica} and full QCD \cite{Fodor:2015doa}, which seem to indicate that the method has problems in the critical region. This concern has been amplified by recent results in low-dimensional strong-coupling QCD \cite{Bloch:2015coa}, where the method was shown to converge to incorrect results for small masses. It is now understood that the CL method only produces correct results if some definite validity conditions are satisfied \cite{Aarts:2011ax,Nagata:2016vkn,Salcedo:2016kyy}, and failure to do this is exactly what goes wrong for certain parameter regions of the theories being investigated.

In this paper, we introduce the reweighted complex Langevin (RCL) method which combines CL and  reweighting to reach regions of parameter space which can otherwise not be reached by the CL method. The principle is simple: we generate a CL trajectory for an auxiliary ensemble with parameter values for which the CL validity conditions are satisfied and then reweight this trajectory of complexified configurations to the target parameter values. One of the upshots of the method is that the validity conditions do not have to be satisfied in the target ensemble.

In contrast to standard reweighting methods, the auxiliary ensemble in this hybrid procedure is taken at parameter values where the action is complex such that it could be closer to the target ensemble, hence making the reweighting more efficient.

The RCL method is generally applicable to theories with complex actions and has now already been tested successfully on a one-dimensional partition function with a strong sign problem \cite{Nishimura:2015pba,Nagata:2016vkn}, random matrix models of QCD \cite{Mollgaard:2013qra,Meisinger2016,Bloch:2016jwt} and QCD in 0+1 and 1+1 dimensions \cite{Bloch:2015coa,Bloch:2017sfg}. 

In Sec.~\ref{sec:cl}, we give a brief introduction of the complex Langevin method and its validity conditions. In Sec.~\ref{sec:rcl}, we introduce the reweighted complex Langevin method, which is the main theme of this paper. In Sec.~\ref{sec:results}, we briefly illustrate the method with results for a one-dimensional partition function and for two-dimensional QCD. Finally, we conclude in Sec.~\ref{sec:conclusions}.

\section{Complex Langevin Method}
\label{sec:cl}

Consider a partition function,
\begin{align}
Z = \int dx \, e^{-S(x)},
\end{align}
with multidimensional real degrees of freedom $x$ and a complex action $S(x)$, which leads to the sign problem.

A Langevin equation based on the complex action is driven by a complex force, which naturally takes the real variables into the complex plane. Therefore, we introduce complexified variables $z = x + i y$, satisfying the CL equation,
\begin{align}
\dot z(t) = -\frac{\partial S}{\partial z} + \eta(t),
\end{align}
with independent multidimensional Gaussian noise $\eta$, which is chosen to be real to get better convergence, satisfying
\begin{align}
\Braket{\eta(t)} &= 0 , \quad
\Braket{\eta(t)\eta(t')} = 2\delta(t-t') .
\end{align}

In practice, the stochastic differential equations need to be discretized, which in the stochastic Euler scheme yields
\begin{align}
z(t+1) = z(t) + \epsilon K + \sqrt{\epsilon}\,\eta ,
\label{Euler}
\end{align}
with drift term $K = -\partial S/\partial z$ and discretized Langevin step size $\epsilon$. 

When applying the complex Langevin method, it is essential that the expectation values computed along the complex trajectory agree with the original expectation values in the partition function with complex action, i.e.,
\begin{align}
\braket{\Obs} \equiv  \int dx \, \rho(x) {\cal O}(x) = \int dx dy \, P(z) {\cal O}(z),
\label{equiv}
\end{align}
where $\rho(x) = e^{-S(x)}/Z$ is the complex weight in the original real variables and $P(z)$ is the real probability in the complex variables $z=x+iy$ along the complex Langevin trajectory.

Proving the equivalence \eqref{equiv} has been the object of thorough study in recent years \cite{Aarts:2011ax,Nagata:2016vkn}, and it is now known to hold if the following validity conditions are satisfied:
\begin{enumerate}[label=(\roman*)]
\item The probability $P(z)$ decays sufficiently rapidly in the imaginary direction of the complexified variables to avoid the \textit{excursion} problem;
\item The probability density $P(z)$ is suppressed close to singularities of the drift and of the observable.
\end{enumerate}
Recently, it was shown that these conditions can be replaced by the single condition that the probability distribution of the drift term should be suppressed, at least exponentially, at large magnitude \cite{Nagata:2016vkn}.

These validity conditions are not always satisfied as was verified for various models and physical systems, in which case the CL method will fail or its expectation values will be incorrect \cite{Mollgaard:2013qra,Makino:2015ooa,Bloch:2015coa,Nishimura:2015pba,Nagata:2016vkn,Salcedo:2016kyy}.

The RCL method proposed below uses reweighting to extend the applicability of the CL method to parameter regions for which the CL validity conditions are not satisfied. 

\section{Reweighting the complex Langevin trajectories}
\label{sec:rcl}

The basic principle of the reweighted complex Langevin method (RCL) is to compute observables for a target ensemble by reweighting the complex CL trajectories generated for an auxiliary ensemble.
For this, it is imperative that the CL validity conditions are satisfied for the auxiliary ensemble, but it is interesting to note that they may be violated in the target ensemble.

We first introduce the standard reweighting method to compute the expectation value of an observable $\Obs$ in a target ensemble characterized by the parameters $\xi$,
\begin{align}
\braket{\Obs}_{\xi} &= \frac{\int dx \, w(x;\xi) \Obs(x;\xi)}{\int dx \, w(x;\xi)} ,
\label{Obs}
\end{align}
with weights $w(x;\xi)= e^{-S(x;\xi)}$ and action $S(x;\xi)$.
To compute this expectation value via reweighting, we also consider an auxiliary ensemble with parameter values $\xi_0$ and rewrite Eq.\ \eqref{Obs} in a completely equivalent form:
\begin{align}
\braket{\Obs}_{\xi} &= \frac{\int dx \, w(x;\xi_0) \left[\Obs(x;\xi)\,\dfrac{w(x;\xi)}{w(x;\xi_0)}\right]}{\int dx \,w(x;\xi_0) \left[\dfrac{w(x;\xi)}{w(x;\xi_0)}\right]}
= \frac{\Braket{\Obs_\xi \dfrac{w_\xi}{w_{\xi_0}}}_{\!\!\xi_0}}{\Braket{\dfrac{w_\xi}{w_{\xi_0}}}_{\!\!\xi_0}} .
\label{rew}
\end{align}
This reweighting equation gives a mathematically exact relation between expectation values in the target ensemble and in an auxiliary ensemble and holds independently of the actions being real or complex. 

In standard practice, the auxiliary ensemble is chosen with real and positive weights so that it can be sampled with importance sampling Monte Carlo methods and the target observables can be estimated as a ratio of sample means according to Eq.\ \eqref{rew}. The target ensemble can have either a real or complex action. An example of real action reweighting is mass reweighting in QCD. However, the more relevant case here is when the target ensemble has complex weights and can itself not be sampled with importance sampling methods, as is the case in QCD at nonzero chemical potential. Reweighting is then one possible way to circumvent the sign problem.

The peculiarity of the new RCL method is that we drop the requirement for the auxiliary weights to be real and positive and allow these to be complex; i.e., we reweight from one ensemble with complex action to another one with complex action. Although Eq.\ \eqref{rew} remains formally correct, this makes a crucial difference from the algorithmic point of view as we can no longer use importance sampling methods to sample relevant configurations in the auxiliary ensemble. Instead, we aim to use the CL method to generate an auxiliary trajectory, i.e., a trajectory in the auxiliary ensemble, along which expectation values in that ensemble can be estimated. For these estimates to be correct, the auxiliary ensemble has to be chosen such that the CL conditions of Sec.~\ref{sec:cl} are satisfied. If we then consider such an auxiliary CL trajectory, the expectation value of any observable $\Obs$ in this ensemble can be computed, according to the CL equivalence \eqref{equiv}, as
\begin{align}
\braket{\Obs}_{\xi_0} = \int dx dy \, P(z;\xi_0) \Obs(z) ,
\label{equivmu}
\end{align}
where $P(z;\xi_0)$ is the real probability in the complex variable along the auxiliary CL trajectory. 

The leap in the RCL method is that we apply the CL formula \eqref{equivmu} to both the numerator and denominator of the reweighting equation \eqref{rew}, hereby specifying the observable $\Obs$ of Eq.~\eqref{equivmu} to the expressions in square brackets in Eq.~\eqref{rew}. This eventually leads to the RCL equation
\begin{align}
\braket{\Obs}_{\xi} 
&= \frac{\int dx dy \, P(z;\xi_0) \left[\Obs(z;\xi)\,\dfrac{w(z;\xi)}{w(z;\xi_0)}\right]}{\int dx dy \, P(z;\xi_0) \left[\dfrac{w(z;\xi)}{w(z;\xi_0)}\right]} .
\label{clrew}
\end{align}
Note that, compared to Eq.\ \eqref{rew}, the expressions in square brackets are now to be evaluated in the complex variable along the auxiliary CL trajectory. In applying Eq.~\eqref{equivmu} to the reweighting equation, an interesting difference between the RCL method and the standard reweighting procedure arises: in the latter, the auxiliary ensemble is sampled according to $w(x;\xi_0)$ and this same factor also occurs in the denominator of the effective observables, whereas in Eq.\ \eqref{clrew} for RCL the ensemble is sampled according to the real probability $P(z;\xi_0)$, but the effective observables contain the complex weights $w(z;\xi_0)$.

An important observation is that, even if the CL validity conditions are not satisfied for the target ensemble, the RCL equation \eqref{clrew} will still yield the correct result as it only uses expectation values in the auxiliary ensemble where the validity conditions are assumed to be satisfied.

In practice, the auxiliary CL trajectory is discretized using  Eq.\ \eqref{Euler}, and Eq.\ \eqref{clrew} is estimated by the ratio of sample means:
\begin{align}
\braket{\Obs}_{\xi}
&\approx \frac{\frac1N \sum_{j=1}^N \Obs(z_j;\xi) \dfrac{w(z_j;\xi)}{w(z_j;\xi_0)}}
{\frac1N \sum_{j=1}^N \dfrac{w(z_j;\xi)}{w(z_j;\xi_0)}} .
\label{RCL}
\end{align}

Although Eq.\ \eqref{rew} is mathematically exact, its practical use is subject to caution. When the auxiliary and target ensembles are not close enough, the accuracy of the method in numerical simulations is plagued by the overlap and sign problems. The overlap problem already occurs when the weights in both ensembles are positive, while the sign problem occurs additionally when the target weights are complex and large cancellations occur in the integrals.

A possible asset of the RCL procedure is that it may allow us to reweight from an auxiliary ensemble that is \textit{closer} to the target ensemble than in standard reweighting, where the auxiliary ensemble needs to have real and positive weights.

\section{Results}
\label{sec:results}

Below we briefly illustrate the RCL method in two examples: a one-dimensional partition function and two-dimensional strong-coupling QCD. 

\begin{figure}
\centerline{\includegraphics[scale=1]{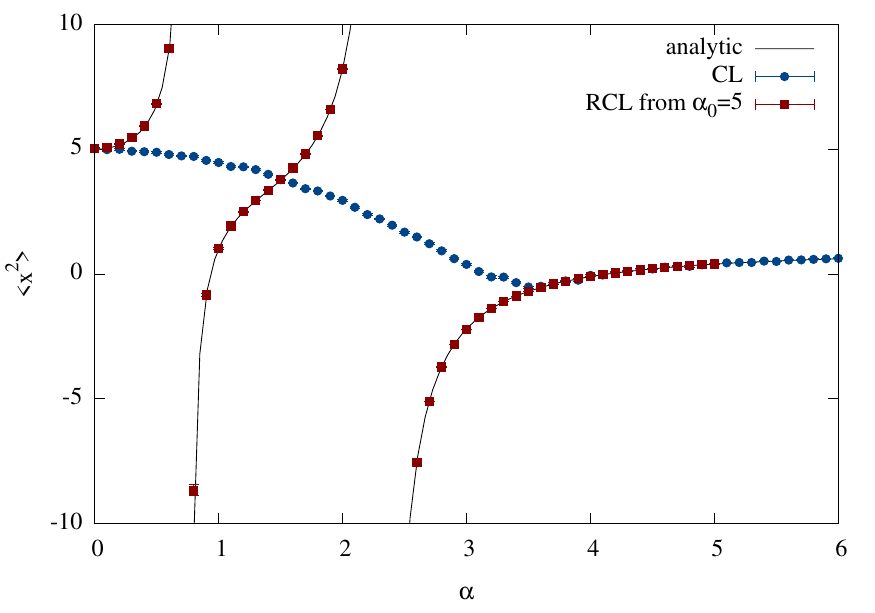}}
\caption{\label{Fig:1d}Results for $\braket{x^2}$ as a function of $\alpha$ for the one-dimensional partition function \eqref{Z1d}. Comparison of CL (blue), RCL (red) and analytical (solid) results.}
\end{figure}

Consider the one-dimensional partition function \cite{Nishimura:2015pba},
\begin{align}
Z = \int_{-\infty}^{+\infty} dx\,(x+i\alpha)^4 e^{-x^2/2} ,
\label{Z1d}
\end{align}
for which $\braket{x^2}$ was shown to converge to a wrong solution for a large parameter range using the CL method, as the trajectories cover the singularity of the drift. In Fig.~\ref{Fig:1d}, we compare the CL and RCL results for $\braket{x^2}$ with the known analytical results. The CL results are wrong when $\alpha\lesssim3.4$; however the RCL, using an auxiliary trajectory generated at $\alpha_0=5.0$, is able to reproduce the correct results over a very large $\alpha$ range, even across the discontinuity jumps in the observable.

As a second illustration we show results for two-dimensional strong-coupling QCD close to the chiral limit. Specific implementation details are given in Ref.\ \cite{Bloch:2017sfg}. The partition function depends on the chemical potential $\mu$ and the quark mass $m$. The sign problem is triggered by the chemical potential: for zero $\mu$, the action is real and there is no sign problem; as $\mu$ is increased the action becomes complex and for a certain range of the parameters, especially for small masses, the sign problem can be very strong. Moreover, the sign problem grows exponentially with the lattice volume. Although CL can, in principle, be used to avoid the sign problem, it was shown in Ref.\ \cite{Bloch:2015coa} that for small masses the CL trajectories cover the singularity of the drift such that the CL validity conditions are violated and the CL results are wrong. 

\begin{figure}
\centerline{\includegraphics[scale=1]{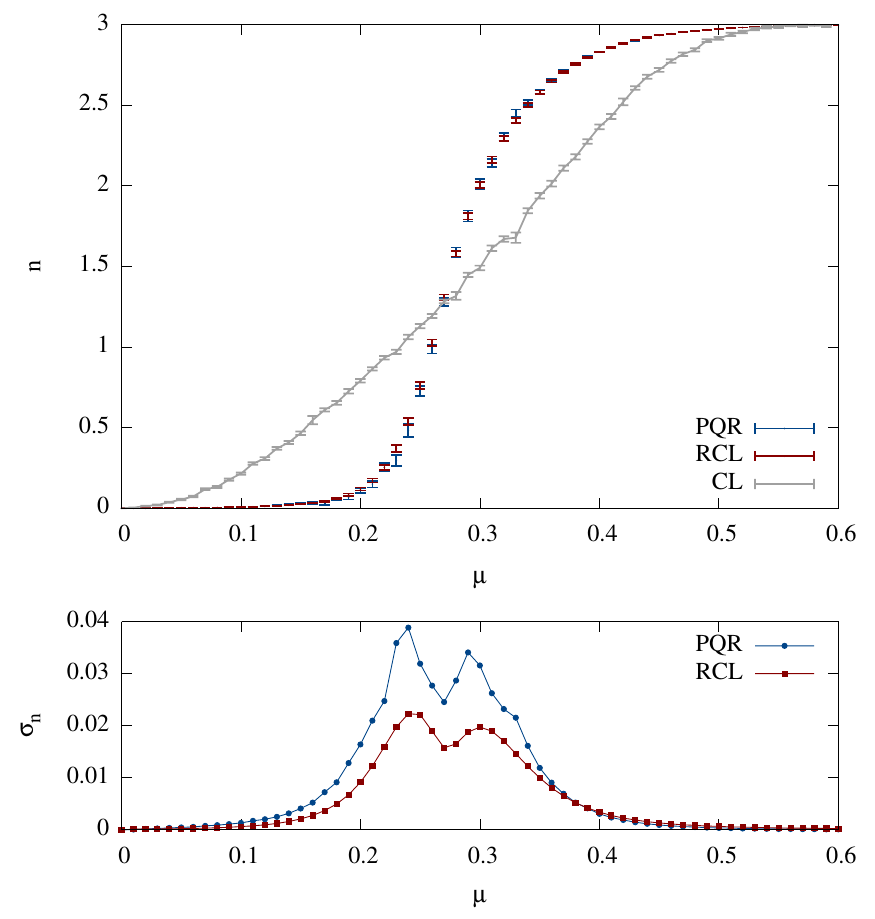}}
\caption{Comparison of RCL and PQR: Quark number density (top) and statistical error (bottom) in two-dimensional QCD as a function of $\mu$ for $\beta=0$ and  $m=0.01$ on a $6\times6$ lattice. The PQR data were computed using Markov chains with 2M configurations; the RCL results using 2M configurations evenly distributed along the auxiliary trajectory at $\mu_0=0.15$ and $m_0=0.1$ containing 10M configurations. For comparison, we also show the wrong CL results.}
\label{Fig:qcd_6x6}
\end{figure}

The failure of the CL method is illustrated in Fig.\ \ref{Fig:qcd_6x6}, where the quark number density is plotted as a function of the chemical potential for a small mass, $m=0.01$, on a $6\times6$ lattice. As a benchmark we use phase-quenched reweighting (PQR) results computed with 2M configurations per $\mu$ value. The CL results clearly disagree with the PQR benchmarks over the complete $\mu$ range. The CL results are, in fact, very similar to those of the phase-quenched theory, as was also observed for a random matrix model for QCD \cite{Bloch:2016jwt}. 

How should we proceed with the RCL simulations? The PQR simulations are expected to encounter problems for $m_\pi/2<\mu<m_N/3$, where the phase-quenched theory is in a pion-condensed phase at low temperature. The aim is to set up the RCL simulations so that the physics of the auxiliary ensemble represents as closely as possible that of the target ensemble. In Ref.\ \cite{Bloch:2017sfg}, the RCL for two-dimensional QCD was performed using an auxiliary CL trajectory at high $\mu$, far above the phase transition. However, at that $\mu$ value, QCD is already in the chirally restored phase and performing RCL from that trajectory may not efficiently capture the phase transition. A better choice may be to perform RCL from a $\mu$-value below the phase transition. However, Fig.\ \ref{Fig:qcd_6x6} shows that there is no valid CL in the low $\mu$-range. Therefore we chose a slightly different strategy and reweight in both the mass and the chemical potential. The auxiliary trajectory is taken from a CL simulation at a somewhat larger mass, $m_0=0.1$, and at $\mu_0=0.15$, which is below the phase transition. This auxiliary point is in the validity region of CL, as was shown in Ref.\ \cite{Bloch:2017sfg}, and can be used to perform RCL. We thus reweight from an auxiliary ensemble at $m_0=0.1$, $\mu_0=0.15$ to target ensembles with $m=0.01$ and a $\mu$ range from 0 to 0.6. 

In practice, we generate a single auxiliary trajectory with 10M configurations and reweight to the 60 target parameter values using one out of every five stored configurations to reduce autocorrelations,\footnote{The remaining autocorrelations are taken care of during the statistical analysis of the RCL measurements.} such that 2M configurations are effectively used in the RCL runs, just as for PQR.  A comparison of the RCL and PQR results for the quark number density is shown in Fig.\ \ref{Fig:qcd_6x6}. The RCL results are in complete agreement with the PQR benchmarks over the entire $\mu$ range. Moreover, as can be seen in the bottom plot, the errors of RCL are almost a factor of 2 smaller than those of PQR in the phase transition region, such that the PQR simulations would require 4 times more CPU time to reach the same level of accuracy.
The time needed to generate the auxiliary trajectory in RCL is easily amortized when computing many target values, as was the case in our run. Note that the RCL data for the various target values are correlated as they originate from a single auxiliary trajectory.

From these results, it seems that indeed the "wrong-phase" problem of PQR can be improved upon by RCL, even though this was only shown in a first preliminary analysis. 
Moreover, it is always useful and even necessary to have independent methods to investigate problems that are far from trivial, as is the case for the sign problem, so that RCL could be a welcome alternative to verify results obtained with PQR.

In both examples, RCL is able to reproduce the correct results, even when CL fails for the target ensemble. In separate publications, we present more detailed RCL results obtained for a random matrix model of QCD \cite{Bloch:2016jwt} and for two-dimensional QCD \cite{Bloch:2017sfg}.

\section{Conclusions}
\label{sec:conclusions}

Previous studies have shown that the complex Langevin method can solve the sign problem occurring in simulations of theories with a complex action; however, caution should be exercised as this requires the CL validity conditions to be satisfied. It turns out that these conditions are often only satisfied for some range of parameters, while they are violated for other parameter values, in which case the CL gives the wrong results. Here we have introduced the \textit{reweighted complex Langevin} (RCL) method, which combines the CL method with a reweighting of the complex trajectories to enlarge the applicability range of the method. As a proof of principle, we presented first results of the method on a one-dimensional partition function and on two-dimensional strong-coupling QCD and verified that the RCL procedure indeed yields the correct results, even when CL itself does not work for the target ensemble. Moreover, we showed that the RCL method can compete with, or even beat, the PQR method if the auxiliary trajectory is chosen in a knowledgeable way.
In the case of two-dimensional QCD, this was done by reweighting both in the mass and in the chemical potential, such that reweighting from below the phase transition was possible.

Although we focused here on cases where CL does not satisfy the validity conditions for the target ensemble, the RCL method can also be applied in cases where CL works fine in the target ensemble with the aim to reuse already existing trajectories obtained at different parameter values and hence gain simulation time. The fact that target and auxiliary ensembles can be chosen close to one another should form a clear advantage over other reweighting methods.

Another possible avenue is to use RCL to interpolate instead of extrapolate in the reweighting parameter, for example the chemical potential in QCD, as we could combine results obtained from auxiliary ensembles at parameter values above and below the target value. This again could improve the quality of the RCL estimates compared to standard reweighting.

Even so, it is to be expected that the overlap and sign problems will deteriorate the efficiency of  RCL, as for any reweighting method, once the auxiliary and target ensembles are too distant. The efficiency of the method and the optimization of reweighting strategies will be studied in more detail in forthcoming work.

\section*{Acknowledgments}
This work was supported by the DFG collaborative research center SFB/TRR-55. Part of the simulations were performed on iDataCool and QPACE~3. I would like to thank R.~Lohmayer, J.~Meisinger, J.~Nishimura, and S.~Schmalzbauer for useful discussions.

\bibliography{biblio} 

\end{document}